\documentstyle[12pt,aasms4]{article}

\begin{document}

\title{Neutrinos and Supermassive Stars: Prospects for Neutrino
Emission and Detection}

\author{Xiangdong Shi and George M. Fuller}
\affil{Department of Physics, University of California, San Diego, La Jolla,
CA 92093}
\authoremail{shi@physics.ucsd.edu, gfuller@ucsd.edu}

\begin{abstract}
We calculate the luminosity and energy spectrum of the neutrino emission
from electron-positron pair annihilation during the collapse of
a supermassive star (${M} \ga 5\times10^4\,{M_\odot}$).
We then estimate the cumulative flux and energy spectrum
of the resulting neutrino background
as a function of the abundance and redshift of supermassive stars and
the efficiency of these objects in converting gravitational energy into
neutrino energy.  We estimate the expected signal in some of the new
generation of astrophysical neutrino detectors from both a
cumulative background of supermassive stars and single collapse events
associated with these objects.
\end{abstract}


\section{Introduction}

In this paper we examine the physics of neutrino emission in the collapse
of supermassive stars.  We also comment on the prospects for future terrestrial
neutrino detectors to obtain a signal from these objects, and we discuss the
possible consequences of such a signal for cosmology.
By \lq\lq supermassive stars\rq\rq\ 
we mean stars so massive that they collapse
on the general relativistic Feynman-Chandrasekhar instability.
This will imply masses for these objects 
${M}\ga 5\times 10^4 {M_\odot}$
(see for example Fuller, Woosley, \& Weaver 1986; thereafter FWW).

Though there is no direct evidence for the existence of these objects,
we note that there is overwhelming evidence for the existence of
supermassive black holes both associated with quasars and Active
Galactic Nuclei (AGN) at high redshift and almost every galaxy-sized
structure examined appropriately by the Hubble Space Telescope (HST)
(van der Marel et al. 1997).
In turn, Begelman and Rees (1978) have shown that a supermassive star could
result as an intermediate stage in the collapse of a relativistic
star cluster to a black hole.

Alternatively, supermassive stars could have formed out of
primordial clouds at high redshifts in which cooling was not
as efficient as in clouds contaminated with metals
(Hoyle \& Fowler 1963; Bond, Arnett, \& Carr 1984;
FWW; McLaughlin \& Fuller 1996).
The typical baryonic Jean's mass at high
redshift can be $\sim 10^5M_\odot$ (Peebles \& Dicke 1968;
Tegmark et al. 1997), but we do not know whether a cloud of this size would
fragment into many pieces and form stars of smaller masses or
collapse directly to form a large object.  
Given the relatively crude understanding we currently possess
on star formation, it is interesting to explore the observational
signatures of supermassive stars, one of which
could be their neutrino emission during core collapse.
(The other telltale signs of the existence and
evolution history of supermassive stars could be nucleosynthesis
products of hot hydrogen burning (the $rp$-process, Wallace
\& Woosley 1981); greatly enhanced local helium and/or deuterium abundances
(Woosley 1977; Fuller \& Shi 1997); or effects of large black holes.)

It would be very significant for our understanding of galaxy formation and
cosmology if there were to be a new neutrino or nucleosynthesis probe of
the epoch of galaxy/quasar formation at redshifts $z\sim$1 to 5.

There are significant differences between neutrino emission in supermassive
stars and in core collapse supernovae.  Neutrinos are trapped and thermalized
in ordinary supernovae.  Because of the different depths inside the
supernova core where the various neutrino species thermally decouple,
the average neutrino energies satisfy the hierarchy
$E_{\nu_\tau}\approx E_{\bar\nu_\tau}\approx
E_{\nu_\mu}\approx E_{\bar\nu_\mu} > E_{\bar\nu_e} > E_{\nu_e}$ .
By contrast, in supermassive stars neutrinos are produced principally
by annihilation of thermal $e^+e^-$-pairs.  They can escape freely
from the core and so their energy spectra are not thermal and
different neutrino species therefore will have similar energy spectra.

The only known limit so far on the neutrino background from
supermassive stars comes from the consideration
that probably no more than $\sim {\cal O}(0.1)$
of all baryons could ever have been in these objects,
otherwise there would probably be too many
$\sim 10^4M_\odot$ to $10^6M_\odot$ relic black holes.
This limit is much less stringent
than the limit on the supernova neutrino background.
This supernova background limit can be obtained because
the past and present supernova rates are subject to a very
tight metallicity production constraint
(Totani, Sato \& Yoshii 1996; Hartmann \& Woosley 1997; Malaney 1997).
The abundance of supermassive stars may not be subject to similar
metallicity concerns because these objects do not necessarily expel
significant amounts of metals into the interstellar medium.
However, while the supernova neutrino background has
contributions from {\it recent} supernovae, the neutrino background
from supermassive stars (if any) probably took its
form at a higher redshift, and therefore suffers a fair amount of
redshift in energy.  Consequently, although the relic neutrino
flux of supermassive stars could be higher than the flux of relic
supernova neutrinos, it has fewer neutrinos with energies in the
range $\ga 10$ MeV.

Numerical calculation of neutrino emissivity from
$e^+e^-$ annihilation have been carried out
previously (Schinder et al. 1987; Itoh et al. 1989).  In this
paper we apply these results to the calculation of
the neutrino luminosity and energy spectra in the
collapse of supermassive stars.  We then proceed to calculate
the flux and energy spectrum of the neutrino background from a
putative population
of supermassive stars as a function of their abundance and redshift, and
their efficiency in converting gravitational energy into neutrino
energy.  We estimate the event rate of this neutrino background
in several new-generation neutrino detectors,
including Super Kamiokande (Super K).
We will also consider the expected event rate of
a single neutrino burst
from the collapse of a single supermassive star.

\section{The Neutrino Luminosity from the Collapse of a Supermassive Star}

Fuller, Woosley, and Weaver (1986) have
discussed the evolution and general relativistic instability of
supermassive stars. In that work it was shown that these objects
will most likely collapse into black holes unless the
centrifugal force resulting from rapid rotation is strong enough
to compensate the build-up of infall kinetic energy.
During the collapse, only part of the star will plunge through the event
horizon and become a \lq\lq prompt\rq\rq\ 
black hole.  This is because the prodigious
thermal neutrino pair emission will render the collapse of a
nonrotating supermassive star non-homologous
(FWW; see also Goldreich \& Weber 1980). 
A simple calculation shows that
\begin{equation}
{M_5^{\rm HC}\over M_5^{\rm init}}\approx
\sqrt{g^{\rm init}\over g^{\rm HC}}
\Bigl({S^{\rm HC}\over S^{\rm init}}\Bigr)^2,
\label{sloss}
\end{equation}
where superscript \lq\lq init\rq\rq\  always refers to quantities in
the initial pre-collapse configuration, and superscript \lq\lq HC\rq\rq\  
refers to quantities of the homologous core during the 
collapse. Here $M_5$ is the stellar mass in units of 10$^5\,M_\odot$;
$g$ is the statistical weight of relativistic particles,
with $g^{\rm init}\approx 2$
and $g^{\rm HC}\approx 2\,+\,(7/8)\times 4=5.5$;
and $S$ is the entropy per baryon.
In the initial nearly isentropic configuration
(assuming an index $n=3$ polytropic structure), we have
\begin{equation}
S^{\rm init} \approx 0.93\Bigl({M^{\rm init}\over M_\odot}\Bigr)^{1/2}-
{4\over\mu},
\label{Sinit}
\end{equation}
where $\mu$ is the mean molecular weight ($\mu\approx 0.59$ for
\lq\lq primordial composition\rq\rq\ 
of 75\% hydrogen and 25\% helium by mass).  In
equation~(\ref{Sinit}) we have assumed that $g=2$ and that all of the
entropy is contributed by photons.  This is an excellent approximation
for a supermassive star near its general relativistic instability point.

As an example, if the
entropy per baryon is reduced by a factor of 2.5
as a result of neutrino emission during the collapse,
i.e., $S^{\rm HC}/S^{\rm init}\approx 0.4$, then the final
homologous core mass will be about 10$\%$ of the initial
stellar mass, $M_5^{\rm HC}/M_5^{\rm init}\approx 0.1$.

The total Newtonian gravitational binding energy of
a homologous core with a mass $M^{\rm HC}_5$ is crudely
$\sim E_{\rm s}\approx 10^{59}M^{\rm HC}_5\,{\rm erg}$.
If there is no strong magnetic field present, most of this energy
will be trapped inside the black hole, radiated through gravitational
waves, or released by the neutrino emission prior to trapped surface
formation.  The characteristic core
radius near the black hole formation point is the
Schwarzschild radius $r_{\rm s}\approx
3\times 10^{10}M^{\rm HC}_5\,{\rm cm}$.  The characteristic duration
of the collapse is the dynamic time $t_{\rm s}
\approx M^{\rm HC}_5\,{\rm sec}$, or longer if rotation or magnetic
fields hold up the collapse.

It has been shown that the neutrino energy loss rate per unit volume,
$Q$, as a result of electron-positron annihilation, has
the simple form (Schinder et al. 1987; Itoh et al. 1989)
\begin{equation}
Q\approx 4\times 10^{15}\,T_9^9\,\,{\rm erg}\,{\rm cm}^{-3}\,{\rm s}^{-3},
\end{equation}
where $T_9\equiv T/10^9$ K.  This equation is valid so
long as the temperature $T\ga m_e$ (the electron rest mass) and
as long as the density $\rho$ is low enough to ensure non-degeneracy
($\rho\la m_pT^3/\hbar^3c^3\sim 10^8(T/1\,{\rm MeV})^3$
${\rm g}\,{\rm cm}^{-3}$ where $m_p$ is the proton rest mass).
Both conditions are satisfied during the collapse of a supermassive star.
The neutrino luminosity from a supermassive star is then
\begin{equation}
L_\nu\approx \epsilon_1\,\epsilon_2\,\int_0^{R} 4\pi r^2\,Q\,{\rm d}r,
\end{equation}
where $R$ is the radius of the homologous core, $\epsilon_1$ is a
factor accounting for the travel time difference for
neutrinos coming out from different depths inside the core, and
$\epsilon_2$ represents the effect of gravitational redshift.

During a homologous collapse, the density $\rho$
has a self-similar profile of polytropic form:
\begin{equation}
\rho=\rho_c\theta^3_3(\xi),
\end{equation}
where $\rho_c$ is the central density of the star, and
$\theta_3$ is the index $n=3$ Lane-Emden function and
$\xi=r/a$ is a dimensionless length measure.  The total pressure
at any point in the star can be cast in the index $n=3$
polytropic form $P=K\rho^{4/3}$.
Here the pressure constant can be expressed as
\begin{equation}
K\approx {1\over 4}\Big({45\over 2\pi^2}\Bigr)
                   \Big({g\over g_s}\Bigr)\,g^{-1/3}\,S^{4/3}
                   \Big[1+{4\over\mu}\Bigl({g_s\over g}\Bigr){1\over S}\Bigr]\,
                   N_A^{4/3},
\end{equation}
where $N_A$ is Avogadro's number and where $g_s$ is the relativistic
particle statistical weight entering into the entropy per
baryon $S\approx (2\pi^2\,g_s\,T^3/45)/\rho\,N_A$.  For all
our considerations in this paper we can safely take $g_s=g$.
If we denote the density in units of $10^3\,{\rm g}\,{\rm cm}^{-3}$
as $\rho_3$, then the pressure in units of MeV$^4$ will be
$P=K_3\rho_3^{4/3}$, with $K_3$ being
\begin{equation}
K_3\approx (6.8\times 10^{-6}\,{\rm MeV}^4)\,\Bigl({11/2\over g_s}\Bigr)^{1/3}
\,S_{100}^{4/3}\,\Big(1+{0.04\over\mu S_{100}}\Bigr)\,
\end{equation}
where $S_{100}$ is the entropy per baryon $S$ in units of 100 Boltzman's
constant.

We can estimate similarly the dimensionless length conversion factor,
\begin{equation}
a={m_{pl}\over \sqrt{\pi}}\,K^{1/2}\,\rho_c^{-1/3}
\approx (8.2\times 10^{10}\,{\rm cm})\,\Bigl({11/2\over g_s}\Bigr)^{1/6}
\,S_{100}^{2/3}\,\Big(1+{0.04\over\mu S_{100}}\Bigr)^{1/2}\,
\Big({\rho_c\over 10^3{\rm g}\,{\rm cm}^{-3}}\Bigr)^{-1/3}
\end{equation}
where, in terms of Newton's constant $G$, the Planck mass is
$m_{pl}=G^{-1/2}$.

Because the entropy per baryon $S_{100}$
is roughly constant through out the radiation-dominated core
at any time, we can estimate that
$T_9^3/\rho_3=0.3S_{100}\approx {\rm constant}$ (FWW).  Therefore,
\begin{equation}
T\approx T_c\,\theta_3(\xi),
\end{equation}
and the quantity 
\begin{equation}
T_9^{\rm aver}\approx (0.3S_{100}\,\bar\rho_3)^{1/3}
\approx (0.3S_{100})^{1/3}\,\Bigl({\int_0^R 4\pi r^2\,\rho_3{\rm d}r\over
\int_0^R 4\pi r^2{\rm d}r}\Bigr)^{1/3}
=(T_c/10^9{\rm K})\Bigl({\int_0^{\xi_1} \xi^2\theta^3_3{\rm d}\xi)
\over (\int_0^{\xi_1} \xi^2{\rm d}\xi}\Bigr)^{1/3}.
\end{equation}
We take $T_9^{\rm aver}$ in
this form for simplicity in calculating the average core density
$\bar\rho$.  The neutrino luminosity can then be expressed through
integrating the Lane-Emden function,
\begin{eqnarray}
L_\nu&\approx (4\times 10^{15}\,{\rm erg}\,{\rm cm}^{-3}\,{\rm s}^{-1})
\,\epsilon_1\,\epsilon_2\,(T_c/10^9{\rm K})^9\,(4\pi a^3)
\int_0^{\xi_1} \xi^2\theta_3^9(\xi){\rm d}\xi \quad &\cr
&=(4\times 10^{15}\,{\rm erg}\,{\rm cm}^{-3}\,{\rm s}^{-1})
\,\epsilon_1\,\epsilon_2\,(T_9^{\rm aver})^9
\Bigl({\int_0^{\xi_1}\xi^2{\rm d}\xi\over
\int_0^{\xi_1} \xi^2\theta_3^3{\rm d}\xi}\Bigr)^3
\,(4\pi a^3)
\int_0^{\xi_1} \theta_3^9(\xi){\rm d}\xi &\cr
&=(1.6\times 10^{18}\,{\rm erg}\,{\rm cm}^{-3}\,{\rm s}^{-1})
\,\epsilon_1\,\epsilon_2\,(T_9^{\rm aver})^9\,(4\pi R^3/3) \quad\quad&
\label{eloss1}
\end{eqnarray}

Apparently most of the neutrinos are emitted near the black hole
formation point, where $T_9^{\rm aver}$ is the highest.
Therefore, to a good approximation we can take the radius of
the core $R$ to be $\beta\,r_{\rm s}
\approx 3\times 10^{10}\beta\,M^{\rm HC}_5\,{\rm cm}$ where
$\beta\ga 1$, with a characteristic dynamic time of
$\beta\,t_{\rm s}\approx \beta\,M^{\rm HC}_5$ sec.
The volume averaged core density is $\bar\rho
\approx M^{\rm HC}/(4\pi\beta^3r_{\rm s}^3/3)=1.83\times 10^6
\,\beta^{-3}\,(M^{\rm HC}_5)^{-2}\,{\rm g}\,{\rm cm}^{-3}$. 
On the other hand, the entropy per baryon in the homologous core (FWW) is
\begin{equation}
S_{100}^{\rm HC}=\alpha\,S_{100}^{\rm init}
=2.9\,\alpha\bigl(M_5^{\rm init}\bigr)^{1/2}
=2.9\,\alpha\bigl(M_5^{\rm HC}\bigr)^{1/2}\,
\bigl({M_5^{\rm init}/M_5^{\rm HC}}\bigr)^{1/2}
\end{equation}
where $\alpha\equiv S_{100}^{\rm HC}/S_{100}^{\rm init}$, the
factor by which the initial entropy is reduced by neutrino emission
in the course of core collapse.  Therefore we conclude that
\begin{equation}
T_9^{\rm aver}=(0.3\,S_{100}\bar\rho_3)^{1/3}
=12\,\alpha^{1/3}\beta^{-1}\Bigl({M_5^{\rm init}\over M_5^{\rm HC}}\Bigr)^{1/6}
(M_5^{\rm HC})^{-1/2}.
\end{equation}
From eq.~(\ref{sloss}) we find
$\alpha^2({M_5^{\rm init}/M_5^{\rm HC}}\Bigr)\approx \sqrt{5.5/2}
\approx 1.66$.  Therefore $T_9^{\rm aver}\approx 13\,
(\beta^2\,M_5^{\rm HC})^{-1/2}$.  As a result, eq.~(\ref{eloss1}) becomes
\begin{equation}
L_\nu\approx 2\times 10^{60}\,\epsilon_1\,\epsilon_2\,\beta^{-6}
(M_5^{\rm HC})^{-1.5}\,{\rm erg}\,{\rm cm}^{-3}\,{\rm s}^{-1}
\label{eloss2}
\end{equation}

An accurate estimate of $\epsilon_1$ and $\epsilon_2$ requires a
3-D numerical treatment with full general relativity.
Given all the other uncertainties inherent
in the problem of supermassive star formation and collapse,
it suffices in this paper to give a \lq\lq ball-park\rq\rq\ 
estimate of these factors using the Newtonian picture.
By assuming that all neutrinos are emitted in the positive radial direction
(not a bad one because most neutrinos are emitted from the
central part of the core, as seen from figure 1),
and that the neutrino emission from
the core is cut off when $R=r_{\rm s}$,
we estimate that $\epsilon_1\approx 1/40$.

Estimating $\epsilon_2$ is trickier.  Because most neutrinos are from
the central region of the core, neutrinos emerging from the core
when the core radius is $R$ are mostly emitted $\sim R/c$ earlier
when the core still has a radius of 
$\sim 2R$ and is less relativistic.  We therefore apply the
core surface redshift factor $\epsilon_2\sim
1-r_{\rm s}/R=1-\beta^{-1}$ (a factor of
$\sqrt{1-r_{\rm s}/R}$ from energy redshift and an additional
$\sqrt{1-r_{\rm s}/R}$ from time dilation to observers) to all neutrinos.  
Although the Newtonian picture simply doesn't apply to
the strong field and time-dependent situation,
we believe that by applying the above $\epsilon_1$ and $\epsilon_2$
to eq.~(\ref{eloss1})
we can get a \lq\lq ball-park\rq\rq\  estimate for the neutrino luminosity
\begin{equation}
L_\nu\sim 5\times 10^{58}\,\beta^{-6}
\,(1-\beta^{-1})\,(M_5^{\rm HC})^{-1.5}\,
{\rm erg}\,{\rm cm}^{-3}\,{\rm s}^{-1}.
\label{eloss3}
\end{equation}

The time evolution of the neutrino luminosity to an observer
at infinity is shown in figure 2, with $\beta\sqrt{1-\beta^{-1}}$
being crudely the observer time axis if
the core radius is collapsing at the speed of light.  The peak luminosity
is reached at $\beta\equiv R/r_{\rm s}\approx 7/6$.  Integrating over 
time yields the total neutrino energy loss up to the
formation of the black hole
\begin{equation}
E_{\nu {\rm loss}}\sim\int_1^{\infty} L_\nu\,\sqrt{1-\beta^{-1}}\,
{\rm d}(\beta t_{\rm s})
\sim 3.6\times 10^{57}\,(M_5^{\rm HC})^{-0.5}\,
{\rm erg}\,{\rm cm}^{-3}\,{\rm s}^{-1}
\approx 0.036\,E_s(M_5^{\rm HC})^{-1.5}.
\label{eloss}
\end{equation}

The energy loss through neutrinos cannot be greater than
the gravitational binding energy itself.  Therefore,
for $M_5^{\rm HC}\la 0.1$
the neutrino loss will saturate the limit of the gravitational binding
energy and the above scaling with $M_5^{\rm HC}$ will break down. 
In fact when $M_5^{\rm HC}\la 0.1$ we do not expect our calculations
to apply in the first place because neutrinos become trapped in the core
instead of freely streaming out.

So far we have assumed that the radius of the
core collapses at the speed of light at the black hole formation point.
The neutrino energy loss and luminosity
can be significant larger than we have calculated
if rapid rotation or strong magnetic fields slow down the collapse
significantly.  In these cases not only will the time available for
neutrino release be longer, but the time delay factor $\epsilon_1$
will be less damaging.

\section{The Neutrino Spectrum from the Collapse of a Supermassive Star}

The energy spectra associated with
neutrino emission from the annihilation of
$e^+e^-$-plasma with a temperature $T$ is best estimated using
a Monte Carlo method: we pick an electron-positron pair from a thermal
distribution at temperature $T$, and randomly distribute the momentum
of one neutrino in all directions in the center-of-mass frame.
The energies and momenta of the two neutrinos produced are then fixed by
momentum and energy conservation.
We then convert the neutrino energies back into the rest frame, and
count each annihilation as
$\vert M\vert^2$ events, where the amplitude of the annilihation matrix
element is (Dicus 1972)
\begin{equation}
\vert M\vert^2={\rm constant}\cdot \{(a^2+b^2)[(p_{e^-}\cdot q_{\nu})
(p_{e^+}\cdot q_{\bar\nu})+(p_{e^-}\cdot q_{\bar\nu})
(p_{e^+}\cdot q_{\nu})]+(a^2-b^2)m_e^2(q_{\nu}\cdot q_{\bar\nu})\}.
\end{equation}
In this equation the $p$'s and $q$'s are four-momenta of $e^+$, $e^-$, $\nu$
and $\bar\nu$, and $m_e$ is the electron rest mass. The coefficients are
\begin{equation}
a=1\pm 2\sin\theta_w,\quad b=0.5,
\end{equation}
where $\sin\theta_w\approx 0.23$, and the $+$ sign
is for $\nu_e$, the $-$ sign is for $\nu_\mu$ and $\nu_\tau$.

After repeating the same process many times (several billion times in our
case), and tallying the total number of events that correspond to various
energy bins, we have an un-normalized neutrino energy spectrum
resulting from $e^+e^-$ annihilation in a steady state equilibrium plasma.
All three neutrino species share a common normalization
so that the calculation gives a relative flux between
$\nu_e\bar\nu_e$ and $\nu_\mu\bar\nu_\mu$
(or $\nu_\tau\bar\nu_\tau$), which is around 4.7:1 as long as
$T\ga m_e\approx 0.5$ MeV.
The simulation also yields the temperature dependence
of the neutrino fluxes, which is proportional to $T^8$ if $T\ga m_e$,
consistent with theoretical expectations.

We {\it fit} the neutrino spectrum
(normalized energy distribution function) to the form
\begin{equation}
f_\nu={1\over T_\nu^3\,F_2(\eta_\nu)}\,{E^2\over e^{(E/T_\nu)-\eta_\nu}+1}.
\label{spectrum}
\end{equation}
where the relativistic Fermi integrals are
\begin{equation}
F_k(\eta_\nu)=\int_0^{\infty}{x^k\,{\rm d}x\over e^{x-\eta_\nu}+1}.
\end{equation}
We find that for $T\ga 0.5$ MeV,
$T_\nu\approx 1.6T$ and $\eta_\nu\approx 2$ for all neutrino species.
The average neutrino energy is $\approx 5.5T$,
higher than that of the ambient $e^+e^-$ plasma.
This is because (1) electrons
and positrons with higher energies have larger cross sections for
annihilation into neutrinos, and (2) the mass of the electron and positron
add into the energy of the neutrinos.

Figure 3 shows the arbitrarily-normalized
neutrino energy spectrum calculated from our Monte Carlo method,
and the analytical fit, at $T=1.5$ MeV.  The analytical fit is remarkably
good.  Similar goodness of fit is also obtained for a variety of
temperatures $T$, ranging
from $0.5$ MeV to $10$ MeV, fully covering the typical temperatures
inside a collapsing supermassive star.

The neutrino spectrum from the collapse of a supermassive star can be
estimated from the single-temperature spectrum eq.~(\ref{spectrum})
averaged over the entire core and over its time evolution. 
A precise account is not warranted at this stage.  As
an estimate, we simply take the $e^+e^-$ temperature to be that at
the peak neutrino emission point in both space and time.  That is,
we estimate it at a position $r\approx 0.14R$ (from figure 1) where
neutrinos emitted at this position emerge from the core
when $\beta\approx 7/6$ (from figure 2).
This turns out to be $T\sim 0.8\,(M_5^{\rm HC})^{-1/2}$ MeV.
Therefore the neutrino spectrum from a supermassive star with a core
mass of $M_5^{\rm HC}$ is to a fair approximation
\begin{equation}
f_\nu\approx {1\over (0.8{\rm MeV}/\sqrt{M_5^{\rm HC}})^3\,F_2(2)}\,
{E^2\over e^{(E\sqrt{M_5^{\rm HC}}/1.2\,{\rm MeV})-2}+1},
\end{equation}
with an average energy
\begin{equation}
\langle E_\nu\rangle\approx 4\,(M_5^{\rm HC})^{-1/2}\,{\rm MeV}.
\label{aver}
\end{equation}
The average energy can be a factor of 2 higher if the collapse is
slowed down significantly by rotation/magnetic fields such that
the travel time difference is much less important.

\section{Neutrino Background from Supermassive Stars and Prospects
for Detection}

If supermassive stars were ever ubiquitous in the universe, they
would have left a significant neutrino background.  Assuming that
a fraction $F$ of all baryons once resided in supermassive stars,
and a fraction $f$ of the associated
gravitational binding energy was released by
neutrino emission, then the total background
flux from this neutrino background is now
\begin{equation}
\phi_\nu\sim Ff\rho_{\rm b}c^3/\langle E_\nu\rangle,
\end{equation}
where $\rho_{\rm b}$ is the
baryon density today, and $\langle E_\nu\rangle$ is the average energy
of the neutrinos from the collapse of supermassive stars.
About 70$\%$ of the flux is $\nu_e\bar\nu_e$, and
15$\%$ of the flux is $\nu_\mu\bar\nu_\mu$ or $\nu_\tau\bar\nu_\tau$.
Since, from eq.~(\ref{eloss}) and (\ref{aver}),
$f_{0.1}=f/0.1$ is usually of order 1, and
$\langle E_\nu\rangle\sim 10$ MeV, we have
\begin{equation}
\phi_\nu\sim 10^5Ff_{0.1}\Bigl({\Omega_{\rm b}h^2\over 0.025}\Bigr)
\Bigl({10{\rm MeV}\over\langle E_\nu\rangle}\Bigr)
\,{\rm cm}^{-2}\,{\rm s}^{-1},
\label{nubkgd}
\end{equation}
where $\Omega_{\rm b}$ is $\rho_{\rm b}$ divided by the critical
density today, and $h$ is the present day Hubble parameter
in units of 100 km/sec/Mpc. 

If $F$ is 10$\%$, a fraction that is certainly allowed
by known constraints, then the flux of background neutrinos from
supermassive stars can be 100 times higher than the neutrino
background from supernovae (Totani, Sato \& Yoshii 1996; Hartmann \&
Woosley 1997; Malaney 1997).
This is because metallicity considerations tightly constrain the
fraction of baryons once in supernova progenitors to be
$\la 10^{-3}$, while the number and energy of neutrinos released 
per baryon may not differ much in the two kinds of collapse event
(supernovae vs. supermassive stars).
The fraction of higher energy neutrinos ($\ga 10$ MeV) in
the neutrino background from supermassive stars is, however, very likely to be
much smaller than that of the supernova neutrino background.
This is because neutrinos coming from supermassive stars suffer
a redshift factor $(1+z)$ if supermassive stars formed and collapsed at a
redshift of $z\ga 1$, making the detection of these neutrinos very difficult.
For example, if $z\ga 3$, the average energy of the neutrino background from
supermassive stars will be only $\la 1.4(M_5^{\rm HC})^{-0.5}$
MeV, far below the threshold of
the currently-running Super Kamiokande experiment (about 7 MeV).
Further complicating the detection of supermassive star
relic neutrinos is the fact that
$\nu_e$ in the background will be hopelessly buried in the solar neutrino flux,
which is $\sim 10^9$ cm$^{-2}\,$s$^{-1}$ at $\approx 1$ MeV
and $\sim 10^6$ cm$^{-2}\,$s$^{-1}$ at $\approx 10$ MeV; and the
$\bar\nu_e$ flux in the background under the optimal condition $F\sim 10\%$
will be comparable to or less than the terrestrial $\bar\nu_e$ background
from nuclear power stations, which ranges from 10$^5$ to $10^7$ cm$^{-2}$
s$^{-1}$ at the various sites of neutrino detectors (Lagage 1985).
Clearly, an earth-bound detection of the neutrino background from supermassive
stars will be extremely difficult, if not impossible.

As worked examples, we calculate the expected event rate from the
supermassive star neutrino background in several
new-generation neutrino detectors, including the currently-running
Super Kamiokande.  The fiducial masses, thresholds and detection channels
of these experiments are listed in table 1.  Figure 4 shows the expected
event rates as a function of $z$, the redshift at which
all supermassive stars existed.  We adopt a neutrino spectrum of
the form in eq.~(\ref{spectrum}) assuming $M_5^{\rm HC}=1$, and we assume
100$\%$ detector efficiencies above the published
thresholds.  These assumptions are not entirely realistic, but they give
good order-of-magnitude estimates.  The event rates scale linearly with
the abundance of supermassive stars (represented by $F$, the fraction
of baryons in these stars), and $f_{0.1}$. 
From figure 4 it can be seen that Super Kamiokande
(Totsuka 1997) may potentially be able to preclude
more than $\sim 10\%$ of baryons ever having been in the
form of supermassive stars at a redshift of $z\ll 1$ (although
such a conjecture may have already been ruled out
by not seeing them directly!).

\section{Neutrino Burst from Collapse of a Single of Supermassive Star}

It is interesting to calculate the neutrino flux from the collapse
of a single supermassive star, and see what the requirement would
be on detectors to observe such an event.
If the collapse occured at redshift $z$, its neutrino fluence now is
\begin{equation}
\phi_\nu\,t\sim 
{fM^{\rm HC}c^2\over 4\pi d_L^2\langle E_\nu\rangle}
\approx (4\times 10^5\,{\rm cm}^{-2})\,f_{0.1}\,(M_5^{\rm HC})^{3/2}
\Bigl({10\,{\rm MeV}\over\langle E_\nu\rangle}\Bigr)\,
\Bigl({6000 {\rm Mpc}\over d_L}\Bigr)^2,
\end{equation}
where $d_L$ the luminosity distance to the star.  The 
average neutrino energy now would be $\langle E_\nu\rangle/(1+z)
\sim 4(M_5^{\rm HC})^{-1/2}(1+z)^{-1}$ MeV.

The duration of the neutrino burst, $t$, is dilated to
$M_5^{\rm c}(1+z)$ sec, or longer if rotation and/or magnetic fields
prolong the collapse.  As long as $z\ga 1$, then we can conclude that
$d_L\sim 6000$ Mpc.  
This burst can be converted roughly into numbers of events in detectors
by scaling from eq.~(\ref{nubkgd}) and figure 4.
The number of events in the detectors are the yearly
event rates in figure 4 (with $F=1$) multiplied by
$10^{-7}f_{0.1}(M_5^{\rm HC})^{3/2}$.
For example, a neutrino burst from a non-rotating non-magnetized
collapsing supermassive star with
$M_5^{\rm c}=1$ at $z=1$ induces $10^{-6}$ event in Super Kamiokande.  This
obviously is impossible to detect.
The $\nu_e$ flux in the burst from the collapse
of one supermassive star when the
$\nu_e$ has a flux and energy comparable to that of the
$^8$B solar neutrinos will be
completely swamped by the solar neutrino flux unless $z\ll 1$.
The $\bar\nu_e$ flux in this example burst, on the other hand,
is comparable to the terrestrial $\bar\nu_e$ background from nuclear reactors.
The supermassive star $\bar\nu_e$ component can only stand out
from the nuclear reactor neutrinos
when $z\ll 1$, when their average energy will be higher than the 3 MeV peak
energy of neutrinos from reactors.  Therefore, it seems that the best
chance to detect such a neutrino burst from a distant supermassive star is
in space, in reactions induced by $\bar\nu_e$, and with extremely
large detectors.  As a simplistic example, to establish a burst signal
requires a minimum of about 10 $\bar\nu_e\,p\rightarrow n\,e^+$ events to be
detected within a duration of $\sim 10$ seconds.  On earth, this requires
a $3\times 10^{10}$ ton Super Kamiokande-type detector
for a supermassive star of $M_5^{\rm HC}=1$ at $z=1$.  But in space, there
may be a significantly lower $\bar\nu_e$
background so that the mass of the detector needed will
be substantially smaller.  One possibility could be an AMANDA
(Antarctic Muon And Neutrino Detector Array, Lowder et al. 1996) with
a threshold of $\la 5$ MeV built on an inactive icy asteroid or
satellite of outer planets.

The frequency of neutrino bursts from supermassive stars (or the
frequency of collapse of supermassive stars) could be non-negligible.
Assuming that they all form and collapse at a redshift $z$, the
frequency of these collapse events as observed now is
\begin{equation}
\sim 4\pi r^2a_z^3{{\rm d}r\over {\rm d}t_0}{\rho_{\rm b}(1+z)^3F\over M},
\end{equation}
where $r$ is the comoving FRW coordinate of these supermassive stars (with
earth at the origin), $a_z$ is the scale factor of the universe at a redshift
$z$ (with $a_0=1$), $t_0$ is the age of the universe,
$\rho_{\rm b}\approx 2\times 10^{-29}{\rm g/cm}^{-2}\,\Omega_{\rm b}h^2
\approx 5\times 10^{-31}{\rm g/cm}^{-2}$ (Tytler \& Burles 1997)
is the baryon density of the universe today, and $M$ is the mass
of a typical supermassive star.  Since ${\rm d}r/{\rm d}t_0=c$, the speed of
light, and $r$ is of order $6000h^{-1}$ Mpc as long as $z>1$,
this frequency is
\begin{equation}
0.3FM_5^{-1}\,{\rm sec}^{-1}\sim 3\times 10^4FM_5^{-1}\,{\rm /day}.
\end{equation}
Therefore, even with $F\sim 0.004\%$ (i.e.,
0.004$\%$ of all baryons were incorporated at one time into 10$^5M_\odot$
supermassive stars), these neutrino bursts would be
occurring on average about once a day,
comparable to the occurrence rate of $\gamma$-ray bursts.
\begin{deluxetable}{cccc}
\footnotesize
\tablecaption{New-generation neutrino detectors for which rates are
calculated.}
\tablewidth{0pt}
\tablehead{\colhead{} & \colhead{Super Kamiokande} & \colhead{SNO}
& \colhead{HELLAZ}}
\startdata
Detection Channel&$\bar\nu_e\,p\rightarrow n\,e^+$
                 &$\nu_e\,d\rightarrow p\,p\,e^-$ (CC)
                 &$\nu\,e^-\rightarrow \nu\,e^-$     \nl
   &$\nu\,e^-\rightarrow \nu\,e^-$
   &$\nu\,d\rightarrow \nu\,p\,n$ (NC)    &     \nl
Threshold        & 7 MeV (current) & 5 MeV & 0.1 MeV     \nl
                 & 5.5 MeV (planned) &     &             \nl
Fiducial Mass    & 22.5 kton & 1 kton & 6 ton   \nl
\enddata
\end{deluxetable}

\section{Summary}

In this paper we have calculated the neutrino luminosity and spectrum from the
$e^+e^-\rightarrow \nu\bar\nu$ process during the collapse of a supermassive
star.  We have estimated that for a typical supermassive star with a homologous
core mass of $10^5M_\odot$, a few percent or higher of the total
gravitational energy can be carried away by neutrinos, with an average
energy $\sim 4$ to 8 MeV, with both the percentage efficiency
and the average energy depending on the timescale of the collapse.
We have further calculated the flux and energy spectrum of the
neutrino background from a population of supermassive stars.  This has been
done as a function of the redshift of these stars, their abundance, and their
efficiency in converting gravitational binding energy into neutrino energy.
We found that the resulting neutrino background could potentially have a
higher flux than the supernova neutrino background.  However,
the average energy of this background flux would likely be lower
because supermassive stars more likely formed at a high redshift,
if they had ever formed at all.
The expected event rates for this background were calculated for several
new-generation neutrino detectors, including Super Kamiokande.
We showed that Super Kamiokande may potentially be able to rule out
the possibility that more than $\sim 10\%$ of baryons could have been
incorporated in supermassive stars
at a redshift $z\ll 1$.  Finally, we calculated the flux and energy
of the neutrino burst resulting from the collapse of a single supermassive
star, and assessed its detectability.  These events are extraordinarily
difficult to detect.  The frequency of such burst events, or
equivalently the collapse event rate of supermassive stars, is also shown to
be possibly significant, easily matching or exceeding the frequency of
occurrence of $\gamma$-ray bursts.

The authors thank Kev Abazajian and Mitesh Patel for helpful discussions.
The work is supported
by NASA grant NAG5-3062 and NSF grant PHY95-03384 at UCSD.

\newpage
\begin{center}
{\Large Figure Caption}
\end{center}
\noindent
Figure 1. Neutrino production distribution inside the homologous core
(un-normalized).  Solid curve: the distribution when
travel time differences are neglected;
dashed line: the distribution when
travel time differences are considered.
\bigskip

\noindent Figure 2. Time evolution of the neutrino luminosity
according to eq.~(\ref{eloss3}), for $M_5^{\rm HC}=1$.
The neutrino emission stops when the radius of the
core $R$ enters the horizon.
\bigskip

\noindent Figure 3. Solid lines: the spectra of the $\nu_e\bar\nu_e$ and
$\nu_\mu\bar\nu_\mu$ (or $\nu_\tau\bar\nu_\tau$) emission due to
$e^+e^-$ annihilation during the collapse of a
supermassive star with an assumed core temperature $T=1.5$ MeV.
The dotted line is the analytical fit to the $\nu_e\bar\nu_e$ spectrum.
A common arbitrary normalization applies to all curves.
\bigskip

\noindent
Figure 4. The event rates from the supermassive star neutrino background
in several new-generation detectors, as a function of the redshift
$z$ of supermassive star collapse, for $M_5^{\rm HC}=1$.

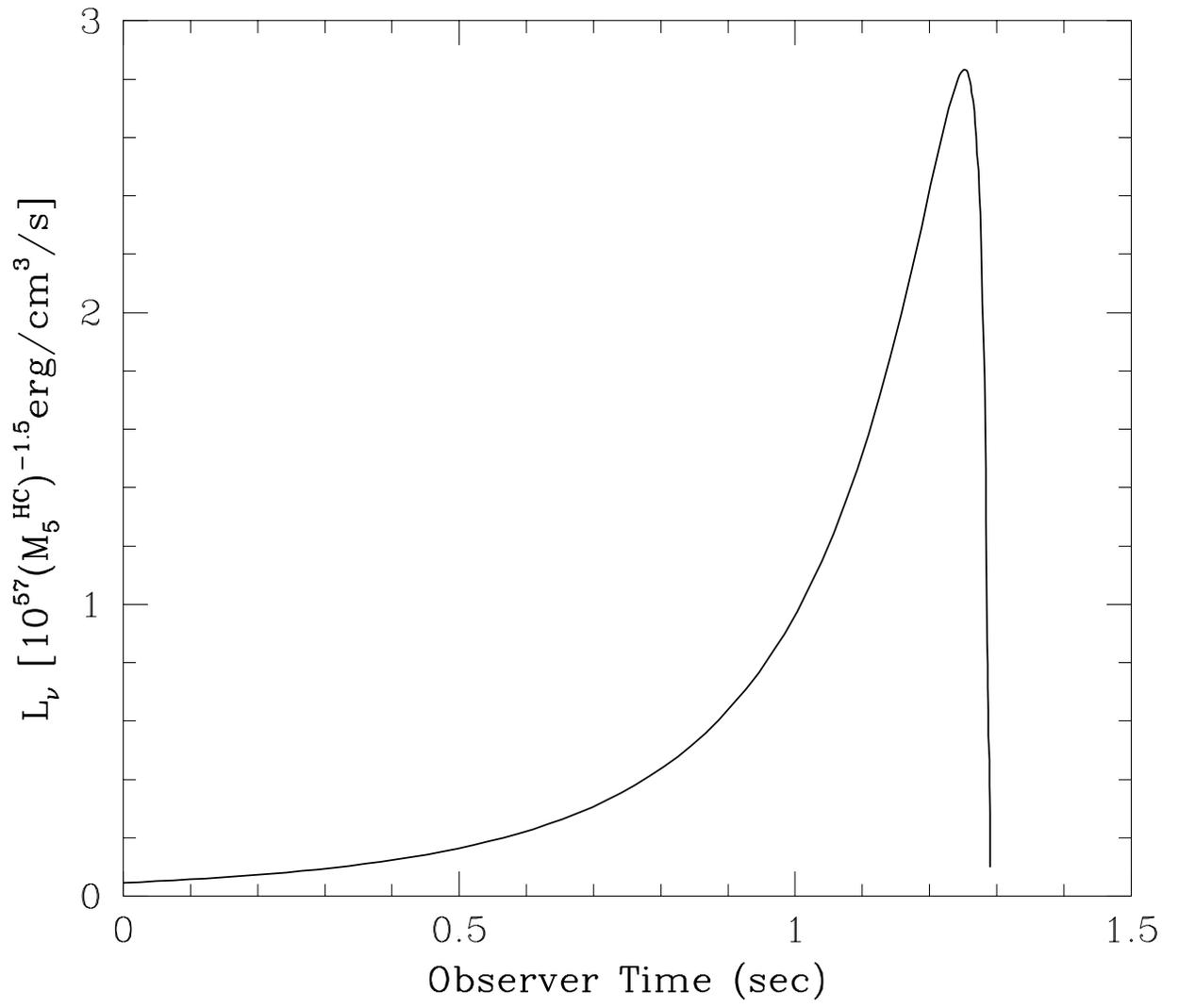
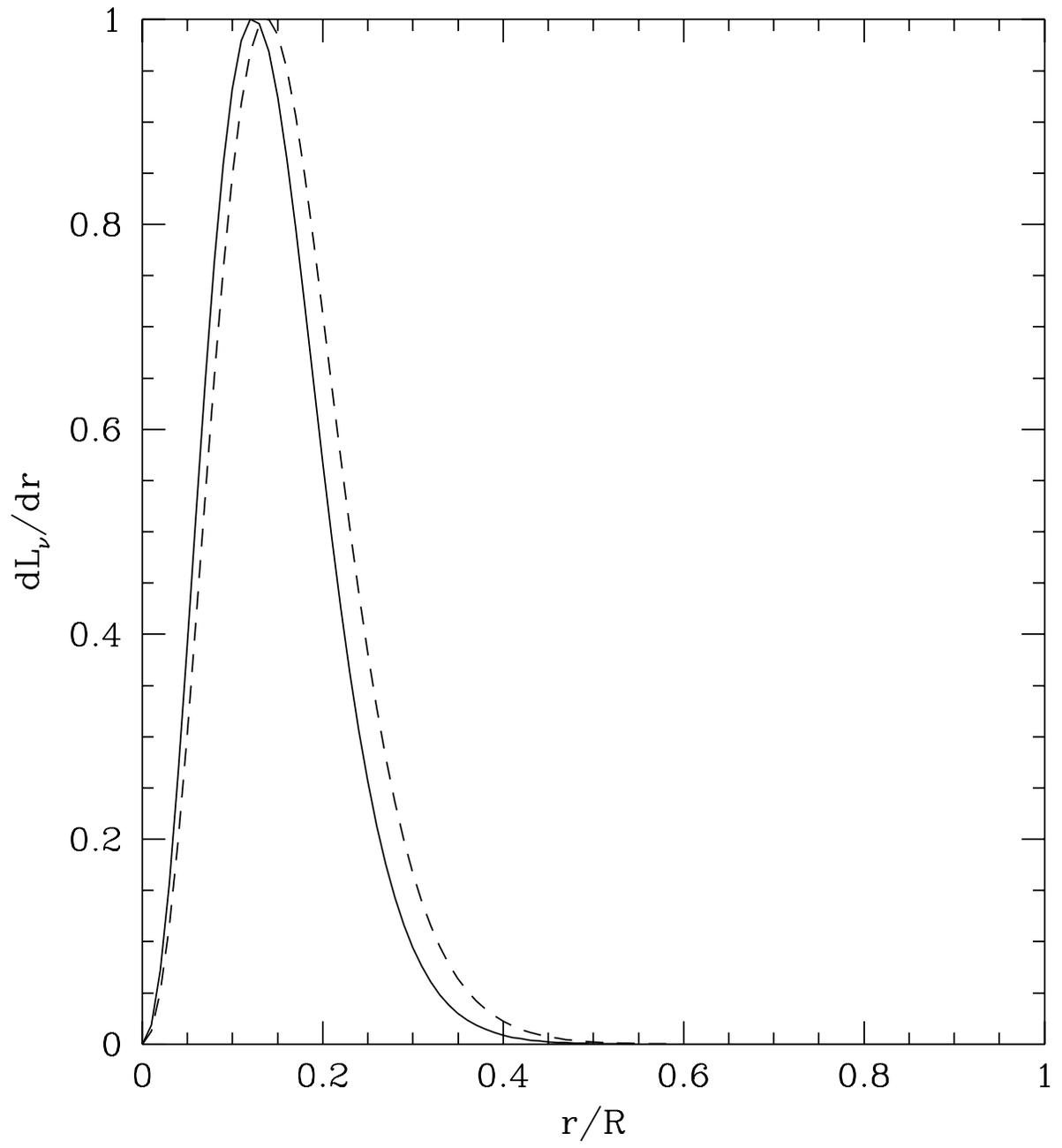
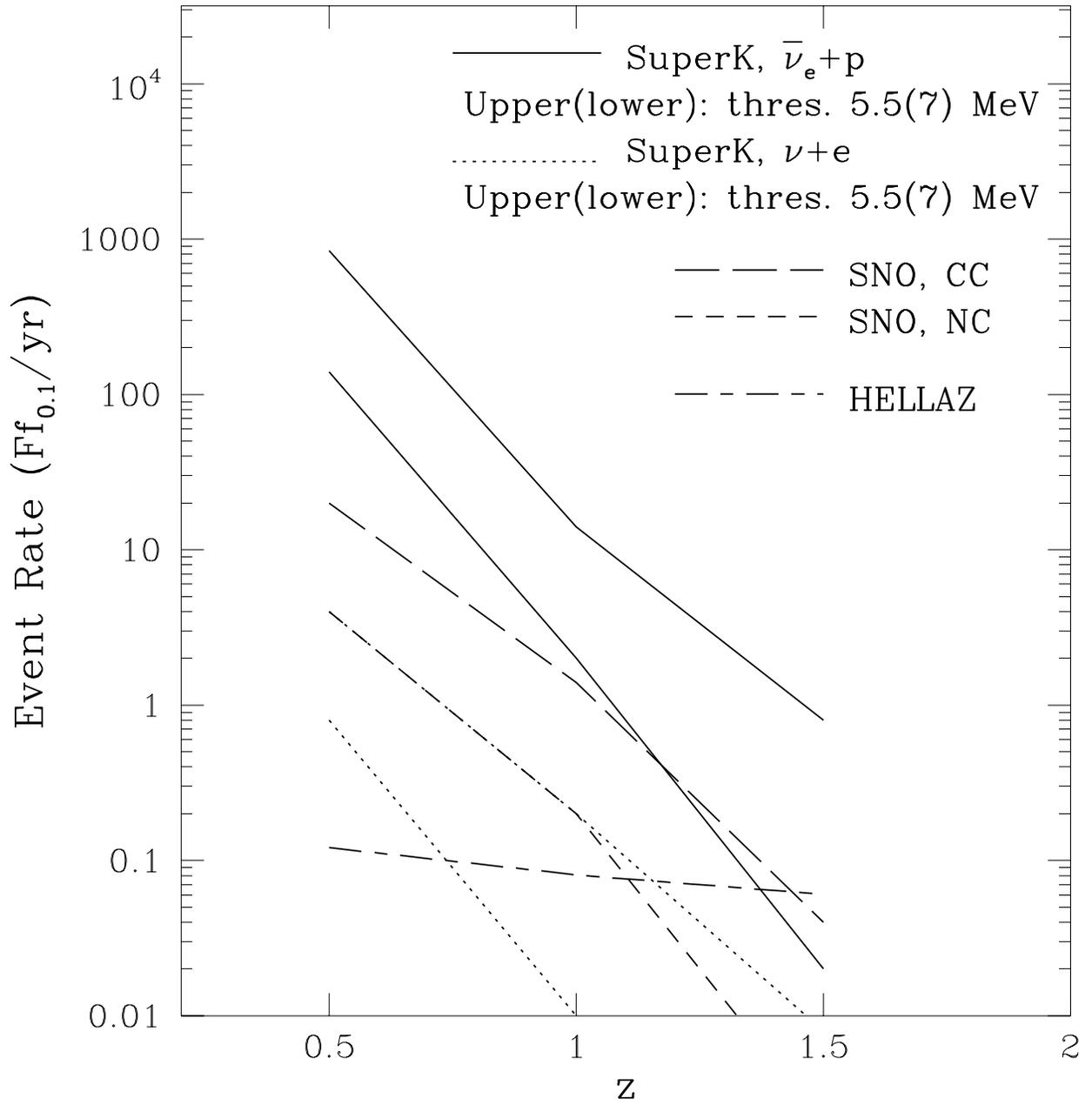
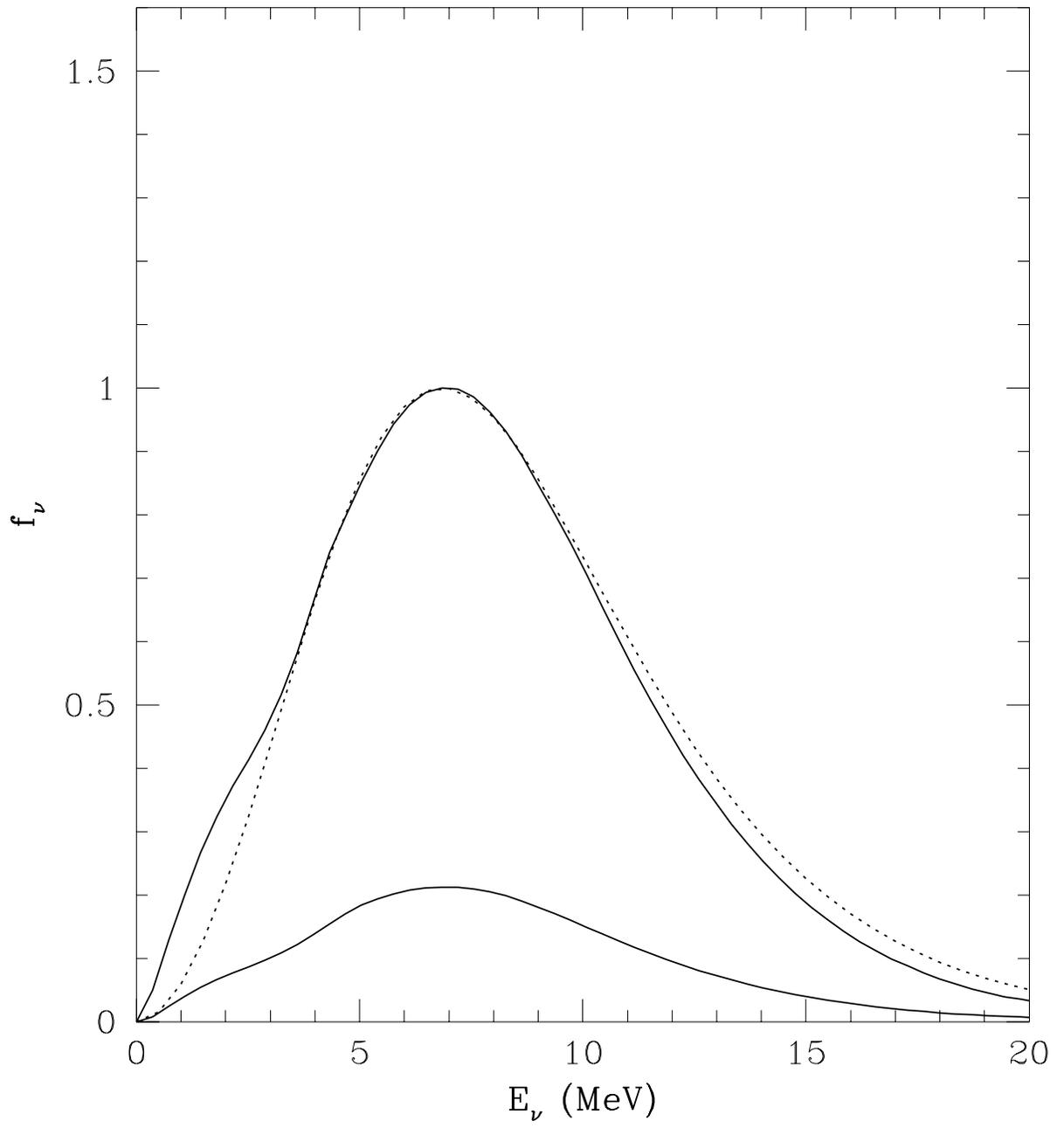
\end{document}